\documentclass[12pt,reqno]{amsart}
\allowdisplaybreaks[4]
\usepackage{latexsym}
\usepackage{amssymb}
\usepackage{amsxtra}
\usepackage{amsmath}
\usepackage{amsfonts}
\usepackage[pdftex]{color}
\usepackage{subfigure}
\usepackage{graphicx}
\usepackage{epstopdf}

\begin{document}
\title{Construction of rational solutions of the real modified Korteweg-de Vries equation from its periodic solutions}
\author{Qiuxia Xing $^1$, Lihong Wang $^1$, Dumitru Mihalache $^2$, \\  Kappuswamy Porsezian $^3$, Jingsong He $^{1,*}$}
\thanks{*Corresponding author: hejingsong@nbu.edu.cn; jshe@ustc.edu.cn}
\dedicatory {1. Department of Mathematics, Ningbo University, Ningbo, Zhejiang 315211, P. R. China\\
2. Horia Hulubei National Institute for Physics and Nuclear
Engineering, \\ P.O. Box MG-6, Magurele, 077125, Romania \\
3. Department of Physics, Pondicherry University, Puducherry 605014,
India
}

\begin{abstract}
In this paper, we consider the real modified Korteweg-de Vries (mKdV) equation and construct a special kind of breather solution, which can be obtained by taking the limit $\lambda_{j}$ $\rightarrow$ $\lambda_{1}$ of the Lax pair eigenvalues used in the $n$-fold Darboux transformation that generates the
order-$n$ periodic solution from a constant seed solution. Further, this special kind of breather solution of order $n$ can be used to generate the order-$n$ rational solution by taking the limit $\lambda_{1}$ $\rightarrow$ $\lambda_{0}$, where $\lambda_{0}$ is a special eigenvalue associated to the eigenfunction $\phi$ of the Lax pair of the mKdV equation. This eigenvalue $\lambda_0$, for which $\phi(\lambda_0)=0$, corresponds to the limit of infinite period of the periodic solution.
Our analytical and numerical results show the effective mechanism of generation of higher-order rational solutions of the mKdV equation from the double eigenvalue degeneration process of multi-periodic solutions.
\end{abstract}

\maketitle \vspace{-0.9cm}

{\bf
During the last 50 years, the concept of solitons has been widely studied
in different branches
of Nonlinear Science and experimentally observed in diverse areas like
hydrodynamics, fiber optics,
quantum gases, Bose-Einstein condensation etc.
Several effective mathematical tools and softwares have been developed to
construct soliton solutions of a
plethora of nonlinear partial differential equations.
During these interesting developments, in addition to soliton solutions,
several other explicit solutions like
dromions, positons, breathers, similaritons, rogue waves etc. have also
been reported for many nonlinear partial
differential equations. In particular, the generation of higher-order
rogue waves from breather-type
periodic solutions has attracted a lot of attention in recent years. To
the best of our knowledge, the concept of breather-positon  is not well investigated. In
this paper, we report breather-positon solutions
to the real modified Korteweg-de Vries (mKdV) equation and construct these
exact solutions from the n-fold Darboux
transformation that generates the order-n periodic solution from a
constant seed solution.
We also generate the order-n rational solutions by taking a suitable limit
in terms
of the eigenvalue of the associated Lax pair of the mKdV equation.
To visualize our above ideas, we consider an optical fiber setting and
demonstrate
that the second limit of double eigenvalue degeneration process might be
realized
approximately by injecting an initial ideal pulse, which is created by a
comb system and a programmable optical filter according to the profile of
the analytical form of the breather-positon at a certain spatial position.
Through this work, we propose  a protocol to observe the higher-order
rational solutions in Kerr-type nonlinear optical media, namely, to measure the wave patterns at the central region
of the higher order breather-positon  generated by ideal initial pulses with suitable limiting
condition.
}

\section{Introduction}

It is a well-known fact that nonlinear partial differential equations play a fundamental role both in the understanding of many natural phenomena and in the development of many new advanced technologies and engineering designs. A plethora of such nonlinear evolution equations have been investigated during the last five decades or so such as the well celebrated Korteweg-de Vries (KdV) equation, the modified Korteweg-de Vries (mKdV) equation, the sine-Gordon (sG) equation, the nonlinear Schr\"{o}dinger (NLS) equation, the Manakov system, the Kadomtsev-Petviashvili equation, the Davey-Stewartson equation, the Maccari system etc. Especially, the origin of the KdV equation and its birth has been a long process and spanned over a period of about sixty years from the initial experiments of Scott-Russell in 1834 \cite{su-1} to the publication in 1895 of a seminal article by Korteweg and de Vries \cite{NK} who developed a mathematical model for the shallow water problem and demonstrated the possibility of solitary wave generation. The KdV equation has been derived from different physical settings, e.g. in plasma physics \cite{E-2,E-1}, hydrodynamics, and in studies of anharmonic (nonlinear) lattices \cite{o-3,o-4}. Here we recall that the existence and uniqueness of solutions of the KdV equation for appropriate initial and boundary conditions have been proved by Sj\"{o}berg \cite{o-5}. It is well known that if $u$ is a solution of the KdV equation $u_{t}+u_{xxx}-6uu_{x}=0$ and $v$ is a solution of the defocusing mKdV equation $v_{t}+v_{xxx}-6v^{2}v_{x}=0$, the two solutions are connected by the Miura transformation \cite{AK1}, namely, $u=v_{x}+v^{2}$. Both KdV and mKdV equations are completely integrable and have infinitely many conserved quantities \cite{AK2}.

The KdV and mKdV equations and their many generalizations were used to describe numerous physical phenomena. For example, the system of coupled KdV equations is a generic model of resonantly coupled internal waves in stratified fluids and can also describe the formation of gap solitons and parametric envelope solitons, see Refs. \cite{B1}-\cite{B5}. In nonlinear optical settings the mKdV equation and its further generalizations were found to adequately describe the ultrashort pulse propagation in nonlinear optical media consisting of only a few optical cycles, beyond the so-called slowly-varying envelope approximation (SVEA) \cite{LS2003}-\cite{M2015}.
The generic mKdV equation adequately describes the propagation of an ultrashort (few-cycle) soliton
in two-level media with the characteristic frequency $\Omega$ that is much larger than the
soliton's characteristic frequency $\omega$ (the so-called long-wave approximation),
see Ref. \cite{LS2003}. On the contrary, when $\Omega$
is much lower than $\omega$ (the so-called short-wave approximation), the propagation of the ultrashort pulses
is described by the sine-Gordon (sG) equation, see Ref. \cite{LS2003}. For two-component
nonlinear optical media, where each component is described by a two-level model,
the combined mKdV-sG equation adequately describes the ultrashort soliton propagation,
see Refs. \cite{L2006,LM2009,SunWu2013}. All these generic equations describe the
propagation of few-cycle pulses beyond the commonly used SVEA, see the review \cite{PhysRep}. We also point out that the generic complex mKdV equation describes the propagation of circularly-polarized few-optical-cycle
solitons in Kerr (cubic) nonlinear media in the long-wave-approximation regime and beyond
the SVEA, see Ref. \cite{L2011}.

Recently, a model based on two coupled mKdV equations was used to describe the soliton propagation in two parallel optical waveguide array, in the presence of linear nondispersing coupling and in the few-cycle regime \cite{Terniche2016}. The mKdV, sG, and mKdV-sG equations were also used in modeling the generation of supercontinuum like white light laser in optical fibers \cite{SCG2014,SCG2016}. The mKdV equation also appears in many other fields of nonlinear science and is responsible for unearthing the underlying science of the nonlinear systems. For example, ion acoustic soliton experiments in plasmas \cite{YOC1,ion1} and fluid mechanics \cite{RW1}, soliton propagation in lattices and acoustic waves in certain anharmonic lattices \cite{H1}, nonlinear Alfv\'{e}n wave propagating in plasma \cite{H2,H22}, meandering ocean
currents  \cite{H3} and the dynamics of traffic flow \cite{on,Y.11}. Furthermore, the mKdV equation is also related to Schottky barrier transmission lines \cite{Y.12}. As a completely integrable dynamical system, the mKdV equation model possesses unique features such as the Painlev\'{e} property \cite{property1,property11}, the Miura transformation \cite{property2}, the inverse scattering transformation \cite{property3}, the Darboux transformation (DT) \cite{4} and so on.

Though the soliton solutions of the KdV and mKdV equations have been widely reported by several papers, due to progress in recent times,  it inspires us to study in the present work other types of soliton solutions, such as a special kind of breather solution of the mKdV equation. To this aim we recall that in Refs. \cite{1,2}, Matveev introduced the concept of a positon as a new solution of the KdV equation,  and then  positon and soliton-positon solutions of the KdV equation were for first time constructed and analyzed. The positons have many interesting properties that differ from those of solitons. The positon is a slowly decaying oscillating solution of a nonlinear completely integrable equation having the special property of being superreflectionless \cite{2}. The positon is weakly localized, in contrast to exponentially decaying soliton solutions. The eigenvalue of the spectral problem is positive (embedded in the continuous spectrum). The positon is completely transparent to other interacting objects. In particular, two positons remain unchanged after mutual collision and during the soliton-positon collision, the soliton remains unchanged, while both the carrier-wave of positon and its envelope experience finite phase-shifts \cite{HE3,onorato}. Thereafter, the positon solutions were constructed for several other models, such as the mKdV equation \cite{positon1992,properties1995}, the sine-Gordon equation \cite{Y.Ohta12}, and the Toda-lattice \cite{YOC}; for an   introductory review on positon theory, see the paper by Matveev \cite{matveevprew}.
 In view of the above interesting properties, it has inspired us to go for further study.  We know that the positon can be obtained from soliton solution by a degeneration process, so it is natural for us to ask whether we can make use of these ideas to construct a special kind of breather solution of the mKdV equation by using a certain Lax eigenvalue degeneration mechanism.

We next consider the focusing real mKdV equation
\begin{equation}\label{MKDV}
\begin{aligned}
q_x=6\alpha q_{ttt}+\alpha q^2q_t,
\end{aligned}
\end{equation}
where
$q=q(x,t)$ is a real function of variables $x$ and $t$, and $\alpha$ is an arbitrary real parameter.
The parameter $\alpha$ can be absorbed in the variable $x$. However, we will still keep it in the above equation in order to be consistent with Ref. \cite{7}.

Considering the above discussion, it is necessary for us to further explore the other possible solutions of the above equation such as rational solutions. In the 1970s, a large number of mathematical efforts were put into the construction of rational solutions of integrable nonlinear partial differential equations. For example, rational solutions were reported for the first time for the celebrated KdV equation. Then, a large number of papers have reported rational solutions of various integrable equations for one dependent function, such as the Boussinesq equation \cite{B}, the Hirota equation \cite{Hirota1}, the Kadomtsev-Petviashvili equation \cite{K} and the NLS equation \cite{NLS1,NLS2,NLS3}; see also the recent works \cite{Re,Rg,Ri}. Furthermore, studies of rational solutions were extended to two coupled integrable systems of nonlinear partial differential equations. A lot of such solutions have been given for vector NLS equations \cite{VNLS1,VNLS2,VNLS3} and for two coupled Hirota equations \cite{Hirota2}. A similar extension has been also reported for three coupled NLS equations \cite{NLS5}. In addition, three fundamental rogue-wave patterns and the key properties of the higher-order rogue waves (i.e., a kind of rational solutions) for the complex mKdV equation have been studied in detail in a recent paper \cite{4}. In general, it is not straightforward to obtain the rational solutions of real nonlinear evolution equations. Although several lower-order rational solutions for the real mKdV equation have been given in Ref. \cite{7}, it is still a challenging task to study the higher-order rational solutions for the real mKdV equation.

We know that the higher-rogue waves can be obtained from multi-breather solutions of the complex mKdV equation, see Ref.  \cite{4}. Two main questions then arise: (1) Can we give the explicit form of the higher order rational solutions ? and (2) Can we give their generating procedure ?
According to those two questions that we have posed, the purpose of this paper is as follows:
\begin{itemize}
\item Construct rational solutions for the real mKdV equation by performing two steps of the eigenvalue degeneration process of order-$n$ periodic solutions (that is, multi-breather solutions);
\item Provide a new and systematic way to generate higher-order rational solutions.
\end{itemize}

In order to realize the above-mentioned double Lax pair eigenvalue degeneration process, we introduce a special kind of periodic solution of the mKdV equation, which we call it a breather-positon solution.  This solution is obtained from the periodic solution in the limit $\lambda_{j}\rightarrow\lambda_{1}$ (here $\lambda_{j}$ are the Lax pair eigenvalues used in the $n$-fold DT, which generates the order-$n$ periodic solution from a constant seed).
Then, the order-$n$ breather-positon solution can be used to generate an order-$n$ rational solution by taking the second limit $\lambda_{1}$ $\rightarrow$ $\lambda_{0}$, where $\lambda_{0}$ is a special eigenvalue associated to the eigenfunction $\phi$ of the Lax pair of the mKdV equation ($\phi(\lambda_0)=0$). The special eigenvalue $\lambda_{0}$ corresponds to the
limit of the period of the periodic solution approaching infinity.
It is worth noting that the above briefly explained mechanism for generating rational solutions of the mKdV equations has been explored in detail in Ref. \cite{5}, for the case of the NLS equation.

The organization of this paper is as follows. In Sec. 2, the order-$n$ periodic solutions of the mKdV equation are reported by using the determinant representation of the DT. In Sec. 3, the explicit form of the order-$1$ (first-order) and order-$2$ (second-order) periodic solutions of the mKdV equation are derived. In Sec. 4, the general form  of order-$n$ breather-positon solution is given and the double eigenvalue degeneration process is studied. In Sec. 5, the construction of rational solutions is analyzed. In Sec. 6, a protocol for possible observation of the rational solutions in Kerr-type nonlinear optical media is briefly discussed. In Sec. 7, the conclusions are made and in Sec. 8 an Appendix is given.

\section{Order-$n$ Darboux transformation of the real mKdV equation}

The Lax pair for the mKdV equation (\ref{MKDV}) has been given \cite{7} as follows:
\begin{equation}\label{1}
\begin{aligned}
\phi_{t}(x,t;\lambda)=M\phi(x,t;\lambda)
\end{aligned}
\end{equation}
\begin{equation}\label{111}
\begin{aligned}
\phi_{x}(x,t;\lambda)=(N_0{ x,t;\lambda}^3+N_1{ x,t;\lambda}^2+N_2{ x,t;\lambda}+N_3)\alpha\phi(\lambda)=N\phi(x,t;\lambda)
\end{aligned}
\end{equation}
with
$$
{\phi(x,t;\lambda)}=\left[\begin{array}{cc}
\phi_{j,1} \\
\phi_{j,2}
\end{array}\right], \qquad
{M}=\left[\begin{array}{ccc}
i\lambda & ir \\
iq & -i\lambda
\end{array}\right],\qquad
{N_{0}}=\left[\begin{array}{ccc}
r_{t}q-q_{t}r & (2r^{2}q+r_{tt})i \\
(2q^{2}r+q_{tt})i & -r_{t}q+q_{t}r
\end{array}\right],
$$
$$
{N_{1}}=\left[\begin{array}{ccc}
2iqr & -2r_{t} \\
2q_{t} & -2iqr
\end{array}\right],\qquad
{N_{2}}=\left[\begin{array}{ccc}
0 & -4ir \\
-4iq & 0
\end{array}\right],\qquad
{N_{3}}=\left[\begin{array}{ccc}
-4i & 0 \\
0 & 4i
\end{array}\right].
$$
Here, $q=r$, $\lambda$ is an eigenvalue parameter, and $\phi$ is the eigenfunction corresponding to the Lax pair eigenvalue $\lambda$. The Eq. (\ref{MKDV}) can be obtained by the zero curvature equation $M_{x}-N_{t}+[M,N]=0$ according to the compatibility condition. It should be noted that, as in Ref. \cite{7}, we will explicitly keep the real parameter $\alpha$ in the above equations in order to compare different kinds of solutions of the real mKdV equation.

In order to preserve the reduction $q=r$ in the $n$-fold Darboux transformation, it is necessary for us to select the eigenfunctions as follows:
$$
\phi_{2j-1}=\phi|_{\lambda=\lambda_{2j-1}}=\left[\begin{array}{cc}
\phi_{2j-1,1}\\
\phi_{2j-1,2}
\end{array}\right]
$$
for the eigenvalue $\lambda_{2j-1}$, and
\begin{equation}\label{3}
\begin{aligned}
\phi_{2j}=\phi{(\lambda_{2j})}=\left[\begin{array}{cc}
\phi_{2j,1}(\lambda_{2j})\\
\phi_{2j,2}(\lambda_{2j})
\end{array}\right]
=
\left[\begin{array}{cc}
-\phi_{2j-1,2}^\ast(\lambda_{2j-1})\\
\phi_{2j-1,1}^\ast(\lambda_{2j-1})
\end{array}\right]
\end{aligned}
\end{equation}
for the
eigenvalues $\lambda_{2j}^{\ast}=\lambda_{2j-1}$, $j=1,2,3,\cdots n$.
We recall that in Ref. \cite{4} it was studied in detail the $n$-fold Darboux transformation that generates a new solution $q^{[n]}$ with determinant representation from a constant seed solution, for the complex mKdV equation.
Similarly, we can obtain here the solution of the real mKdV equation as follows:
\begin{equation}\label{3a}
\begin{aligned}
q^{[n]}=q^{[0]}+2\frac{N_{2n}}{D_{2n}},
\end{aligned}
\end{equation}
where $q^{[0]}$ is a seed solution, and
$$
{N_{2n}}=\left[\begin{array}{ccccccc}
\phi_{11} & \phi_{12} &\lambda_{1}\phi_{11}& \lambda_{1}\phi_{12}& \cdots\lambda_{1}^{n-1}\phi_{11}& \lambda_{1}^{n}\phi_{11}\\
\phi_{21} & \phi_{22} &\lambda_{2}\phi_{21}& \lambda_{2}\phi_{22}& \cdots\lambda_{2}^{n-1}\phi_{21}& \lambda_{2}^{n}\phi_{21}\\
\phi_{31} & \phi_{32} &\lambda_{3}\phi_{31}& \lambda_{3}\phi_{32}& \cdots\lambda_{3}^{n-1}\phi_{31}& \lambda_{3}^{n}\phi_{31}\\
\vdots & \vdots &\vdots & \vdots & \vdots & \vdots\\
\phi_{2n,1} & \phi_{2n,2} &\lambda_{2n}\phi_{2n,1}& \lambda_{2n}\phi_{2n,2}& \lambda_{2n}^{n-1}\phi_{2n,1}& \lambda_{2n}^{n}\phi_{2n,1}
\end{array}\right],
$$
$${W_{2n}}=\left[\begin{array}{ccccccc}
\phi_{11} & \phi_{12} &\lambda_{1}\phi_{11}& \lambda_{1}\phi_{12}& \cdots\lambda_{1}^{n-1}\phi_{11}& \lambda_{1}^{n-1}\phi_{12}\\
\phi_{21} & \phi_{22} &\lambda_{2}\phi_{21}& \lambda_{2}\phi_{22}& \cdots\lambda_{2}^{n-1}\phi_{21}& \lambda_{2}^{n-1}\phi_{22}\\
\phi_{31} & \phi_{32} &\lambda_{3}\phi_{31}& \lambda_{3}\phi_{32}& \cdots\lambda_{3}^{n-1}\phi_{31}& \lambda_{3}^{n-1}\phi_{32}\\
\vdots & \vdots &\vdots & \vdots & \vdots & \vdots\\
\phi_{2n,1} & \phi_{2n,2} &\lambda_{2n}\phi_{2n,1}& \lambda_{2n}\phi_{2n,2}& \lambda_{2n}^{n-1}\phi_{2n,1}& \lambda_{2n}^{n-1}\phi_{2n,2}
\end{array}\right].$$
Note that we only keep the symbol $``,"$ in the last line of the above expression. There are $2n$ real parameters in $\lambda_{j}=R_{0j}+iR_{j}$$(j=1,3,5,\cdots, 2n-1)$ and two real variables $x$ and $t$ associated with the eigenfunctions  in the explicit expression of $q^{[n]}$. We will next set $R_{0j}=0$ in order to simplify our calculations.

\section{The first-order and second-order periodic solutions of the mKdV equation}

We start with a special constant seed solution $q^{[0]}=1$. The eigenfunction $\phi$ is obtained for the corresponding eigenvalue by precise mathematical operations as follows
\begin{equation}\label{2}
\begin{aligned}
\phi(\lambda)=\left[\begin{array}{cc}
d_{1}(\lambda)e^{cti-2ic(2\lambda^2-1)\alpha x}-d_{2}(\lambda)e^{-cti+2ic(2\lambda^2-1)\alpha x}\\
d_{1}(\lambda)\frac{ie^{cti-2ic(2\lambda^2-1)\alpha x}}{\lambda i+ci}-d_{2}(\lambda)\frac{ie^{-cti+2ic(2\lambda^2-1)\alpha x}}{\lambda i-ci}
\end{array}\right],
\end{aligned}
\end{equation}
where $d_{1}{(\lambda)}=e^{icS}$, $d_{2}{(\lambda)}=e^{-icS}$, $S=S_{0}+\sum_{k=0}^{n-1}{s_{k}\epsilon^{2k}}$, $c=\sqrt{\lambda^2+1}$.
We should notice that the parameters $s_{k}$ that were introduced above are crucial to adjust the phase of the breather, and that  $\epsilon$ is also an important parameter, which is used to acquire the degeneration limit of eigenvalues by using a Taylor expansion for constructing the breather-positon solutions and then the corresponding rational solutions. In order to simplify the following tedious calculations, we will set the eigenvalue $\lambda$ as a pure imaginary number.

According to the expression of $\phi(\lambda)$ given in Eq. (\ref{2}), an explicit form of eigenfunction $\phi_{2j-1} (j=1,3, 5,\dots, n)$ is given by
\begin{equation}\label{explicitphi2j-1}
\begin{aligned}
\phi_{2j-1}=\phi(\lambda_{2j-1})=\left[\begin{array}{cc}
d_{1}(\lambda_{2j-1})e^{cti-2ic(2\lambda_{2j-1}^2-1)\alpha x}-d_{2}(\lambda_{2j-1})e^{-cti+2ic(2\lambda_{2j-1}^2-1)\alpha x}\\
d_{1}(\lambda_{2j-1})\frac{ie^{cti-2ic(2\lambda_{2j-1}^2-1)\alpha x}}{\lambda_{2j-1} i+ci}-d_{2}(\lambda_{2j-1})\frac{ie^{-cti+2ic(2\lambda_{2j-1}^2-1)\alpha x}}{\lambda_{2j-1} i-ci}
\end{array}\right].
\end{aligned}
\end{equation}
Meanwhile, the $\phi_{2j}$ is constructed from $\phi_{2j-1}$ by using the reduction conditions in Eq. (\ref{3}),
i.e.
\begin{equation}\label{explicitphi2j}
\begin{aligned}
\phi_{2j}
=
\left[\begin{array}{cc}
-\phi_{2j-1,2}^\ast(\lambda_{2j-1})\\
\phi_{2j-1,1}^\ast(\lambda_{2j-1})
\end{array}\right],
\end{aligned}
\end{equation}
which is associated with eigenvalue $\lambda_{2j}=\lambda_{2j-1}^*$. Note that $\lambda_0=i$ is a zero of the eigenfunction $\phi$, i.e., $\phi(\lambda_0)=0$, which implies that the period of the breather solution goes to infinity, and thus the breather becomes the rational solution of the mKdV equation. This fact is very crucial to generate higher order rational solutions later by higher order Taylor expansion in determinants with respect $\epsilon$, through $\lambda_j= \lambda_0+ \epsilon$.
Substituting the above eigenfunctions associated with the seed $q^{[0]}=1$ back into Eq. (\ref{3}), then it yields order-$n$ periodic (breather) solutions $q^{[n]}_{\rm br}$. For example, setting $n=1$, we get
\begin{equation}\label{4}
\begin{aligned}
q^{[1]}_{\rm br}=q^{[0]}+2\frac{N_{2}}{W_{2}},
\end{aligned}
\end{equation}
with
$$
{N_{2}}=\left|\begin{array}{ccc}
\phi_{11}& \lambda_1\phi_{11}\\
\phi_{21} & \lambda_2\phi_{21}
\end{array}\right|,
\qquad
{W_{2}}=\left|\begin{array}{ccc}
\phi_{11} & \phi_{12} \\
\phi_{21} & \phi_{22}
\end{array}\right|.
$$
After tedious simplifications we obtain
\begin{equation}\label{5}
\begin{aligned}
q^{[1]}_{\rm br}=-1+2\frac{R_{1}^2-1}{\sin(\frac{2ab}{3})R_{1}+\cos(\frac{2ab}{3})R_{1}^2-1},
\end{aligned}
\end{equation}
where $\lambda_{1}=iR_{1}$, $a=\sqrt{-R_{1}^2+1}$, and $b=2R_{1}^2x-3s_{0}-3t+x$. It is suggested that the order-$1$ periodic solution $q^{[1]}$ is nonsingular. We know that the periodic solution is obtained by the Darboux transformation, and the denominator is $\phi_{11}^2+\phi_{12}^2$. According to Eq. (\ref{2}), we can easily find that $\phi_{12}\neq0$ when $\phi_{11}=0$. Further, it is worth noting that similar periodic solution have been obtained in Ref. \cite{7}.

Fig. 1(a) shows the order-$1$ periodic solution whose amplitude remains constant, and the distance between two peaks is always the same. We see in Fig. 1 the occurring of the typical parallel line waves; the waveforms plotted in this figure are quite different from those corresponding to order-$1$ breather solutions of the complex mKdV equation, see Ref. \cite{4}.
The main difference is the highest values of the wave field are located on spots instead of being on parallel lines. According to Ref. \cite{4}, those spots can be approximated by lines when  $a\rightarrow0$ in $q^{[1]}$. It is easy to note that the spots become lines when $a=0$, and $|q^{[1]}|^{2}$ is a soliton propagating along a line $x=6c^2t$. For the real mKdV equation,  $q^{[1]}_{br}$ gives the exact expression for a soliton propagating along the line $t=-6\alpha x$. From the keen observation of Fig. 1 $(a)$, $(c)$, and $(e)$, it is easy to find that the distance between neighbouring two peaks is larger when $R$ is  closer to $1$. At the same time, the number of peaks decreases from four to three, and further to a single one. By comparing Fig. 1 $(e)$ and Fig. 4 $(a)$, we can find that when the value of $R_{1}\rightarrow1$,
the order-$1$ periodic (breather) solution is very close to the order-$1$ rational solution; see the details in the next sections.

Similarly, we can get the order-$2$ periodic breather solution $q^{[2]}_{\rm br}$ form Eq. (\ref{3}) by setting the eigenvalues  $\lambda_{2}=\lambda_{1}^{\ast}$ and $\lambda_{4}=\lambda_{3}^{\ast}$, where the eigenfunctions are defined by Eqs.  (\ref{explicitphi2j-1}) and (\ref{explicitphi2j}). This solution is expressed by
\begin{equation}\label{6}
\begin{aligned}
q^{[2]}_{\rm br}=q^{[0]}+2\frac{N_{4}}{W_{4}},
\end{aligned}
\end{equation}
with
$$
{N_{4}}=\left|\begin{array}{ccccc}
\phi_{11} & \phi_{12} &\lambda_{1}\phi_{11}& \lambda_{1}^2\phi_{11}\\
\phi_{21} & \phi_{22} &\lambda_{2}\phi_{21}& \lambda_{2}^2\phi_{21}\\
\phi_{31} & \phi_{32} &\lambda_{3}\phi_{31}& \lambda_{3}^2\phi_{31}\\
\phi_{41} & \phi_{42} &\lambda_{4}\phi_{41}& \lambda_{4}^2\phi_{41}
\end{array}\right|, \qquad
{W_{4}}=\left|\begin{array}{ccccc}
\phi_{11} & \phi_{12} &\lambda_{1}\phi_{11}& \lambda_{1}\phi_{12}\\
\phi_{21} & \phi_{22} &\lambda_{2}\phi_{21}& \lambda_{2}\phi_{22}\\
\phi_{31} & \phi_{32} &\lambda_{3}\phi_{31}& \lambda_{3}\phi_{32}\\
\phi_{41} & \phi_{42} &\lambda_{4}\phi_{41}& \lambda_{4}\phi_{42}
\end{array}\right|.$$
The explicit expression of order-$2$ periodic solution $q^{[2]}_{\rm br}$ can also be obtained. This solution will be presented in Appendix, and it is plotted in Fig. \ref{fig.2} for different values of the parameters. It is well-known that the order-$2$ periodic solution is the nonlinear superposition of two periodic solutions that cross each other, which is the main reason why  there are some raised peaks for each wave train in Figs. \ref{fig.2} (a), (c), and (e). Thus, the order-$2$ periodic solution $q^{[2]}_{\rm br}$ actually creates a sort of a two-dimensional lattice structure. The highest peaks appear at the intersection of troughs of one periodic wave structure with the maxima of the every other one.

It is easy to see that the denominator in the expression of the order-$2$ periodic solution $q^{[2]}_{\rm br}$ is zero in the degenerate case when $\lambda_{1}=\lambda_{3}$. In general, the order-$n$ periodic solution becomes an indeterminate form $\frac{0}{0}$ when $\lambda_{j}\rightarrow\lambda_{1} (j=1,3,5,\cdots 2n-1)$. In the next Section, we will study the degenerate limit of the Lax pair eigenvalues corresponding to order-$n$ periodic solutions.

\section{The  breather-positon solution of the mKdV equation}

We know that the order-$n$ rational solutions can be generated by a double degeneration mechanism $\lambda_{j}\rightarrow\lambda_{1}$ and $\lambda_{1}\rightarrow\lambda_{0}$ for the case of the NLS equation, see Ref.  \cite{5}.
But the double degeneration process can be realized in a single step as $\lambda_{j}\rightarrow\lambda_{0}$, by using a Taylor expansion technique. The positon solution can be obtained by the degeneration of soliton solution with zero seed solution, $q=0$, which is clearly stated in earlier Matveev's papers \cite{1}-\cite{2}. Based on these papers \cite{1}-\cite{2}, one can define the degeneration of multi-soliton solutions for the KdV and mKdV equations. So when the seed solution is $q=1$, the periodic solution obtained by this degeneration mechanism is called a breather-positon solution, which can be expressed as $q^{[n]}_{\rm b-p}$ in the limit of $\lambda_{j}\rightarrow\lambda_{1}$. Specifically, the eigenvalue $\lambda_{1}\neq\lambda_{0}$. Further, it can be defined $\lambda_{2j+1}\rightarrow\lambda_{1}$ and $\lambda_{2j}\rightarrow\lambda^{\ast}_{1}$ according to $\lambda_{2j}=\lambda^{\ast}_{2j-1}$.
According to the definitions that we have given, we will study the computing method of the breather-positon solution for the real mKdV equation. We also notice that the breather-positon solution of NLS equation has been studied in Ref. \cite{4}, where the order-$n$ breather-positon was obtained by a Taylor expansion of $q^{[n+1]}$ in the limit $\lambda_{j}\rightarrow\lambda_{1}$.
Similarly, taking the eigenfunctions given by Eqs. (\ref{explicitphi2j-1}) and (\ref{explicitphi2j}) back into Eq. (\ref{3}), and doing higher-order Taylor expansion in $q^{[n]}$ through $\lambda_j=\lambda_1+\epsilon$, then the order-$n$ breather-positon solution
of the mKdV equation is obtained as
\begin{equation}\label{7}
\begin{aligned}
q^{[n]}_{\rm b-p}=q^{[0]}+2\frac{N_{2n}'}{W_{2n}'},
\end{aligned}
\end{equation}
where
$$
N_{2n}'=(\frac{\partial^{n_{i}-1}}{\partial\epsilon^{n_{i}-1}}|_{\epsilon=0}(N_{2n})_{ij}(\lambda_{1}+\epsilon))_{2n\times2n},
$$
$$
W_{2n}'=(\frac{\partial^{n_{i}-1}}{\partial\epsilon^{n_{i}-1}}|_{\epsilon=0}(W_{2n})_{ij}(\lambda_{1}+\epsilon))_{2n\times2n},
$$
$n_{i}=[\frac{i+1}{2}]$, ${[i]}$ defines the floor function of $i$, and $q^{[0]}=1$. It is rather easy to find that an order-$1$ breather-positon solution is an order-$1$ periodic solution, which is given in Eq. (\ref{5}). The first nontrivial breather-positon solution is the order-$2$ breather-positon $q^{[2]}_{\rm b-p}$, which is the limit of an order-$2$ breather in the limit $\lambda_3\rightarrow \lambda_1$. This limit is visually demonstrated in Fig. \ref{fig.2} when $R_2$ goes to $R_1=0.5$. The explicit form of $q^{[2]}_{\rm b-p}$ is provided in Appendix; in Fig. \ref{fig.3} we see the gradual process of approaching the order-$2$ rational solution from an order-$2$ breather-positon solution.
It is useful to demonstrate intuitively the two limits of double degeneration mechanism in a graphical way based on analytical solutions $q^{[2]}_{br}$ and $q^{[2]}_{\rm b-p}$, i.e. the transition of an order-$2$ periodic breather solution to an order-$2$ breather-positon solution in the limit $\lambda_3\rightarrow \lambda_1$, and then the transition of an order-$2$ breather-positon to an order-$2$ rational solution in the limit $\lambda_1\rightarrow \lambda_0$.
\begin{itemize}
\item Fig. \ref{fig.2} shows that the number of the wave trains of periodic solution gradually decreases when $\lambda_3\rightarrow \lambda_1$ until a single wave train is preserved, when an order-$2$ periodic solution becomes an order-$2$ breather-positon solution.
\item Fig. \ref{fig.3} shows that the peaks around the central region of the breather-positon waveform gradually shift until only the central field profile survives and all other accompanying peaks disappear, implying that the order-$2$ breather-positon becomes the order-$2$ rational solution.
\end{itemize}
By looking at Figs. \ref{fig.2} (e) and (f) and Figs. \ref{fig.3} (a) and (b), we see that in order to emphasize the double degeneration process $\lambda_3\rightarrow \lambda_1\rightarrow\lambda_0$ we used the same set of parameters ${\alpha, R_{1}}$.

\section{The rational solution of the mKdV equation}

\begin{figure}[!htbp]
\centering
\subfigure[$s_0=0$, $R_{1}=0.4$]{\includegraphics[height=6cm,width=6cm]{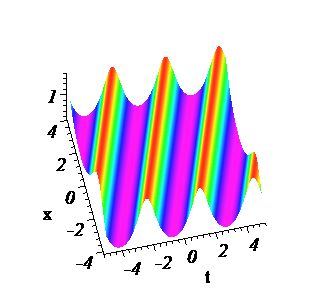}}\qquad\qquad
\subfigure[$s_0=0$, $R_{1}=0.4$]{\includegraphics[height=5cm,width=5cm]{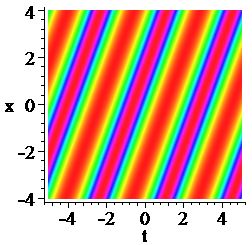}}
\subfigure[$s_0=0$, $R_{1}=0.9$]{\includegraphics[height=6cm,width=6cm]{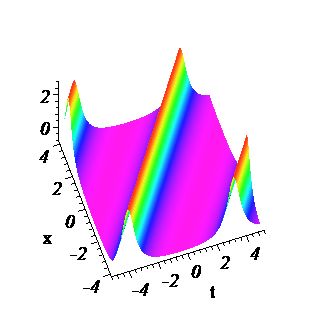}}\qquad\qquad
\subfigure[$s_0=0$, $R_{1}=0.9$]{\includegraphics[height=5cm,width=5cm]{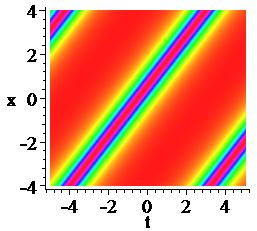}}
\subfigure[$s_0=0$, $R_{1}=0.96$]{\includegraphics[height=6cm,width=6cm]{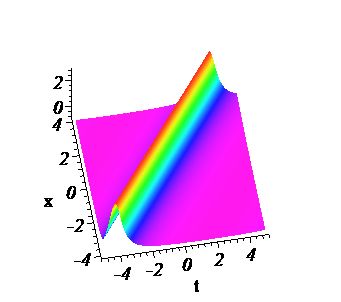}}\qquad\qquad
\subfigure[$s_0=0$, $R_{1}=0.96$]{\includegraphics[height=5cm,width=5cm]{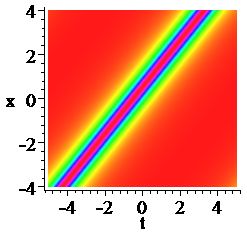}}
\caption{The order-$1$ breather $q^{[1]}_{\rm br}$ gradually approaches the rational solution
 when $R_{1}$ goes to $1$. The right column is the density plot of the left column. The last row is almost the same as the
profile of the order-$1$ rational solution in Fig. \ref{fig.4}. Here $\alpha=-\frac{1}{6}$. }\label{fig.1}
\end{figure}

The order-$n$ rational solutions of the mKdV equation is obtained by setting $\lambda_1\rightarrow \lambda_0$ in Eq. (\ref{7}),
\begin{equation} \label{rationaln}
q^{[n]}_{\rm r}=q^{[n]}_{\rm b-p}(\lambda_1=\lambda_0+\epsilon)��
\end{equation}
This limit is realized by a higher-order Taylor expansion. The first-order breather-positon solution is an order-$1$ periodic solution in Eq.  (\ref{5}), which generates an order-$1$ rational solution
\begin{equation}\label{9}
\begin{aligned}
q^{[1]}_{\rm r}=-1+\frac{2}{(6\alpha x+t)^2+(6\alpha x+t+1)^2}.
\end{aligned}
\end{equation}
The above solution is obtained from order-$1$ periodic solution by Taylor expansion in $\epsilon$, where  $\lambda_{1}=\lambda_{0}+\epsilon$. Figure \ref{fig.1} provides an intuitive idea of the generation process of rational solution  starting from a single-breather solution (that is, from an order-$1$ breather-positon solution) and approaching an order-$1$ rational solution. Here, we set $d_{1}=d_{2}=1$ in order to simplify our calculations. From Fig. \ref{fig.1} and Fig. \ref{fig.4}, it can be observed that the order-$1$ periodic solution is similar to order-$1$ rational solution when $\lambda_{1}$ is very close to $\lambda_{0}$, but $\lambda_{1}\neq\lambda_{0}$. Note that such rational solution was recently given in Ref. \cite{7}.
Similarly, setting $n=2$ and using a Taylor expansion in Eq. (\ref{rationaln}), the order-$2$ breather-positon solution becomes an order-$2$ rational solution. We clearly see this transition process by looking at Fig. \ref{fig.3}(e) and Fig. \ref{fig.5}(a).

The explicit form of the order-$2$ rational solution is given as follows
\begin{equation}\label{10}
\begin{aligned}
q^{[2]}_{\rm r}=1+\frac{F^{[2]}_{\rm r}}{G^{[2]}_{\rm r}},
\end{aligned}
\end{equation}
where
\begin{eqnarray*}
\begin{aligned}
F^{[2]}_{\rm r}=&-62208\alpha^4x^4+(-41472\alpha^3t-20736\alpha^3)x^3+(-10368\alpha^2t^2-10368\alpha^2t-12096\alpha^2)x^2\\
&+(-1152\alpha t^3-1728\alpha t^2-2880\alpha t-1008\alpha)x-48t^4-96t^3-144t^2-72t,
\end{aligned}
\end{eqnarray*}
\begin{eqnarray*}
\begin{aligned}
G^{[2]}_{\rm r}=&746496\alpha^6x^6+(746496\alpha^5t+373248\alpha^5)x^5+(311040\alpha^4t^2+311040\alpha^4t+10368\alpha^4)x^4\\
&+(69120\alpha^3t^3+103680\alpha^3t^2+20736\alpha^3t-5184\alpha^3)x^3+(8640\alpha^2t^4+17280\alpha^2t^3\\
&+8640\alpha^2t^2+864\alpha^2t+4896\alpha^2)x^2+(576\alpha t^5+1440\alpha t^4+1344\alpha t^3+720\alpha t^2+864\alpha t\\
&+216\alpha)x +16t^6+48t^5+72t^4+72t^3+72t^2+36t+9.
\end{aligned}
\end{eqnarray*}
The limit of $\lambda_1\rightarrow \lambda_0=i$ is demonstrated visually in Fig. \ref{fig.3} when $R_1$ goes to $1$. Thus the processes of double eigenvalue degeneration from an order-$2$ breather to an order-$2$ rational solution are clearly shown in Fig. \ref{fig.2} and Fig. \ref{fig.3}, respectively. Our results support the claim that a higher-order rational solution is indeed generated from a multi-breather solution via a double eigenvalue degeneration mechanism.

We should note that the order-$2$ rational solution and order-$3$ rational solution of the mKdV equation (\ref{MKDV}) are much different from the corresponding ones for the complex mKdV equation, which were reported in Ref. \cite{4}.
It is easy to observe that for the order-$2$ rational solution of the mKdV equation there is a single peak at $(x=-1/8, t=-5/8)$, see Fig. \ref{fig.5}.
We see from Fig. \ref{fig.5} that the order-$2$ rational solution of the mKdV equation is a combination of two different types of waveforms: a bright one and a dark (dip) one, a pattern that is much different from that obtained in the case of the complex mKdV equation; see Fig. 1 in Ref. \cite{4}.
Also, note  that the order-$3$ rational solution has two characteristic bright peaks, see Fig. \ref{fig.6}.
Also, according to Ref. \cite{4}, for specific sets of parameters, the order-$2$ rogue wave of complex mKdV equation can be completely separated into three order-$1$ rogue waves arranged in a triangular pattern (see the left panel in Fig. 8 of Ref. \cite{4}). Similarly, the order-$3$ rogue wave of complex mKdV equation can be completely separated into six order-$1$ rogue waves arranged in a triangular pattern (see the right panel in Fig. 8 of Ref. \cite{4}).
However, the rational solutions of the real mKdV equation cannot be separated in terms of order-$1$ rational solutions.
Similar to the case of the order-$2$ rational solution of the mKdV equation (see Eq. (\ref{10})), the order-$3$ and order-$4$ rational solutions of the mKdV equation can be explicitly obtained by using the analytical formula given in Eq. (\ref{rationaln}); these solutions are plotted in Fig. \ref{fig.6} and Fig. \ref{fig.7}, respectively. An explicit form of the order-$3$ rational solution is given in Appendix.

\section{A protocol for possible observation of the rational solutions in optical media with Kerr-type (cubic) nonlinearity}

It is quite clear from Fig. \ref{fig.3} that the conversion process of a breather-positon into a rational solution is quite similar to the conversion of an order-$1$ periodic solution to an order-$1$ rational solution, and the later process can be observed in Fig. \ref{fig.1} and Fig. \ref{fig.4}. Note that an adequate experimental technique was used to observe an order-$1$ rogue wave of the NLS equation in optical fiber settings \cite{10,11,aa}. Further, the typical pattern of the breather-positon solution in its central region provides a good approximation of the corresponding rational solution. Thus, in principle, the characteristic features of the breather-positon solutions might be used to observe the higher-order rational solutions in physical settings involving specially engineered Kerr-type (cubic) nonlinear media.
Based on the two key features of the breather-positon solution, i.e., its convenient conversion to the rational solution and the easy availability of the wave pattern in the central region of the ($x,t$)-plane, we advance the following protocol to observe higher-order rational solutions:
\begin{itemize}
\item Select suitable values of the parameters $s_i$, $\lambda_1$, and $\lambda_0$ to generate the typical breather-position wave pattern. Next, select a suitable position $x_0$, and then obtain the ideal initial pulse $q(x_0, t)$ of the breather-positon;
\item Use a frequency comb and a wave shaper to create the above ideal initial pulse $q(x_0,t)$, and then inject it into a suitable Kerr-type (cubic) nonlinear optical medium;
\item Measure the  values of output pulses $q$  at one or several positions $x_1, x_2,
x_3, \cdots$, which are functions of $t$ and are  denoted by $q_1, q_2, q_3, \cdots$, and then compare them with the  theoretical curves of analytical breather-positon solutions, i.e., with $q(x_1,t)$,
$q(x_2,t)$, $q(x_3,t)$, etc., in order to confirm the expected agreement between theoretical predictions and experimental values.
\end{itemize}

In Fig. \ref{fig.8}, panel (a), we plot a typical input pulse, and the other two panels give the shapes of the output pulse at two different positions, in order to confirm the statement that when the parameter $R_{1}$ is very close to $1$, the order-$2$ breather-positon soluton is an excellent approximation of the order-$2$ rational solution; see also  Fig. \ref{fig.9}.
Note that the corresponding pulses at the same spatial locations are very similar with respect to each other in Figs. \ref{fig.8} and \ref{fig.9}, strongly supporting the statement that the breather-positon is indeed an excellent approximation of the corresponding rational solution of the real mKdV equation.
We think that the above findings might be demonstrated in physical settings involving Kerr-type nonlinear media, such as optical fiber systems and other specially engineered nonlinear materials, e.g., in physical settings illustrated in Fig. \ref{fig.1} of Ref. \cite{aa} and in Fig. \ref{fig.3} of Ref. \cite{bb}.

\section{Summary and discussion}

In this paper, the order-$n$ periodic solutions of the real mKdV equation are expressed in terms of the determinant representation of the corresponding Darboux transformation. Then, we introduce a new special kind of periodic solution, called breather-positon solution, which can be obtained by taking the limit $\lambda_{j}$ $\rightarrow$ $\lambda_{1}$ of the Lax pair eigenvalues used in the $n$-fold Darboux transformation that generates the order-$n$ periodic solution from a constant seed solution.
We have also provided an explicit formula for the breather-positon solution by using the determinant representation of the Darboux transformation and the higher-order Taylor expansion in Eq. (\ref{7}). Further, the order-$n$ breather-positon solution can be converted into a order-$n$ rational solution by performing the limit $\lambda_{1}\rightarrow\lambda_{0}$.  Here  $\lambda_{0}$ is a special eigenvalue associated with the eigenfunction $\phi$ of the Lax pair of the mKdV equation.
According to analytical formulas we derived in this paper, we have illustrated graphically in Fig. \ref{fig.2} the dynamics of the transition from the order-$2$ breather to the order-$2$ breather-positon solution. In Fig. \ref{fig.3} we have illustrated graphically the subsequent transition of the order-$2$ breather-positon solution to the order-$2$ rational solution.
The two main advantages of the breather-positon solution, namely, its convenient conversion into the rational solution and the easy controllability of the wave patterns in the central region of the $(x,t)$ plane, suggested us to propose a protocol to observe higher-order rational solutions in physical settings involving Kerr-type nonlinear optical media.
In conclusion, we have put forward via a systematic approach, a generating mechanism of higher-order rational solutions of the real modified Korteweg-de Vries equation from a double eigenvalue degeneration process of multi-periodic solutions.

\section{APPENDIX}

Because of the tedious expression of $q^{[2]}_{\rm br}$, we only present its final form with the explicit values $s_{k}=0, R_{2}=4/5, R_{1}=3/5, \alpha=-1/6 $ as follows:
\begin{equation}\label{11}
\begin{aligned}
q^{[2]}_{\rm br}=1+\frac{F^{[2]}}{G^{[2]}},
\end{aligned}
\end{equation}
where
\begin{eqnarray*}
\begin{aligned}
F^{[2]}=&2688\sin f_{2}-1512\sin f_{4}-3584\cos f_{2}+1134 \cos f_{4}+2450,\\
G^{[2]}=&-150\sin f_{1}-2100\sin f_{2}-2058\sin f_{3}+2100\sin f_{4}+2800\cos f_{2}+7056 \cos f_{3} \\
&-1575 \cos f_{4}-8425,
\end{aligned}
\end{eqnarray*}
and
$
f_{1}=\frac{14t}{5}-\frac{686x}{375}, f_{2}=\frac{6t}{5}-\frac{114x}{125}, f_{3}=\frac{2t}{5}-\frac{2x}{375}, f_{4}=\frac{8t}{5}-\frac{344x}{375}.
$
The order-$2$ breather-positon expression $q^{[2]}_{\rm b-p}$ is explicitly obtained when $R_{1}=\frac{4}{5}$ as follows:
\begin{equation}\label{12}
\begin{aligned}
q^{[2]}_{\rm b-p}=-1-\frac{F^{[2]}_{\rm b-p}}{G^{[2]}_{\rm b-p}},
\end{aligned}
\end{equation}
where
\begin{eqnarray*}
\begin{aligned}
F^{[2]}_{\rm b-p}=&(-602112\sin g_{3}\cos g_{3}-677376\cos g_{3}^2+338688\cos g_{2}+301056\sin g_{2}-564480)x^2\\
&+(4300800\sin g_{3}\cos g_{3}t+4838400\cos g_{3}^2t+168000\sin g_{3}\cos g_{3}+112000\cos g_{3}^2\\
&-860160\cos g_{3}\sin g_{4}+250880\cos g_{3}\cos g_{4}-2419200\cos g_{2}t-2150400\sin g_{2}t\\
&-2761920\cos g_{2}-3094560\sin g_{2}+430080\sin g_{1}-125440\cos g_{1}+4032000t+3976000)x\\
&-7680000\sin g_{3}\cos g_{3}t^2-8640000\cos g_{3}^2t^2-600000\sin g_{3}\cos g_{3}t-400000\cos g_{3}^2t\\
&+3072000\cos g_{3}\sin g_{4}t-896000\cos g_{3}\cos g_{4}t+4320000t^2\cos g_{2}+3840000\sin g_{2}t^2\\
&+9864000\cos g_{2}t+11052000\sin g_{2}t-1536000\sin g_{1}t+448000\cos g_{1}t-7200000t^2\\
&+12000000\sin g_{2}-3840000\sin g_{1}+1120000\cos g_{1})-14200000t-2031250,
\end{aligned}
\end{eqnarray*}

\begin{eqnarray*}
\begin{aligned}
G^{[2]}_{\rm b-p}=&(-200704\cos g_{3}^2\cos g_{2}+150528\cos g_{3}^2\sin g_{2}+451584\sin g_{3}\cos g_{3}+338688\cos g_{3}^2\\
&-150528\cos g_{3}\sin g_{4}+200704\cos g_{3}\cos g_{4}-169344\cos g_{2}-225792\sin g_{2}+37632\sin g_{1}\\
&-50176\cos g_{1}+332416)x^2+(1433600\cos g_{3}^2\cos g_{2}t-1075200\cos g_{3}^2\sin g_{2}t\\
&-3225600\sin g_{3}\cos g_{3}t-2419200\cos g_{3}^2t+1075200\cos g_{3}\sin g_{4}t-1433600\cos g_{3}\cos g_{4}t\\
&-4116000\sin g_{3}\cos g_{3}-112000\cos g_{3}^2+2016000\cos g_{3}\sin g_{4}+112000\cos g_{3}\cos g_{4}\\
&+1209600\cos g_{2}t+1612800\sin g_{2}t-268800\sin g_{1}t+358400\cos g_{1}t+2016000\cos g_{2}\\
&+3738000\sin g_{2}-1008000\sin g_{1})-56000\cos g_{1}-2374400t-1960000)x+10015625\\
&+7000000t-2560000\cos g_{3}^2\cos g_{2}t^2-7200000\cos g_{3}\sin g_{4}t+2560000\cos g_{3}\cos g_{4}t^2\\
&-400000\cos g_{3}\cos g_{4}t+1920000\sin g_{2}\cos g_{3}^2t^2+5760000\sin g_{3}\cos g_{3}t^2\\
&-1920000\cos g_{3}\sin g_{4}t^2+14700000\sin g_{3}\cos g_{3}t+4240000t^2-9000000\cos g_{2}\\
&+1920000\sin g_{1}-2625000\sin g_{2}-560000\cos g_{1}-2160000t^2\cos g_{2}-13350000\sin g_{2}t\\
&+3600000\sin g_{1}t-640000\cos g_{1}t^2+200000\cos g_{1}t-7200000\cos g_{2}t+4320000\cos g_{3}^2t^2\\
&-2880000\sin g_{2}t^2+480000\sin g_{1}t^2+400000\cos g_{3}^2t,
\end{aligned}
\end{eqnarray*}
and
$g_{1}=\frac{12t}{5}-\frac{228x}{125}, g_{2}=\frac{6t}{5}-\frac{114x}{125}, g_{3}=\frac{3t}{5}-{\frac{57x}{125}}, g_{4}=\frac{9t}{5}-\frac{171x}{125}.
$
By using the Taylor expansion for $\lambda_{1}=\lambda_{0}+\epsilon$, the order-$3$ breather-positon solution generates the following order-$3$ rational solution:
\begin{equation}\label{10a}
\begin{aligned}
q^{[3]}_{\rm r}=-1-\frac{F^{[3]}_r}{G^{[3]}_r},
\end{aligned}
\end{equation}
where
\begin{eqnarray*}
\begin{aligned}
F^{[3]}_{\rm r}=&-3072x^{10}+(30720t+15360)x^9+(-138240t^2-138240t-46080)x^8
+(368640t^3+552960t^2\\
&+368640t+92160)x^7
+(-645120t^4-1290240t^3-1290240t^2-645120t-714240)x^6\\
&+(774144t^5+1935360t^4+2580480t^3+1935360t^2+3271680t+1235520)x^5+(-645120t^6\\
&-1935360t^5-3225600t^4-3225600t^3-6105600t^2-4219200t+254400)x^4+(368640t^7\\
&+1290240t^6+2580480t^5+3225600t^4+5990400t^3+5443200t^2+134400t-379200)x^3\\
&+(-138240t^8-552960t^7-1290240t^6-1935360t^5-3340800t^4-3369600t^3-547200t^2\\
&+331200t+28800)x^2+(30720t^9+138240t^8+368640t^7+645120t^6+1059840t^5+1108800t^4\\
&+288000t^3-216000t^2-259200t-86400)x-3072t^{10}-15360t^9-46080t^8-92160t^7\\
&-161280t^6-198720t^5-129600t^4-43200t^3-4050,
\end{aligned}
\end{eqnarray*}

\begin{eqnarray*}
\begin{aligned}
G^{[3]}_{\rm r}=&512x^{12}+(-6144t-3072)x^{11}+(33792t^2+33792t-1024)x^{10}+(-112640t^3-168960t^2\\
&+26880)x^9+(253440t^4+506880t^3+46080t^2-195840t-26880)x^8+(-405504t^5\\
&-1013760t^4-245760t^3+599040t^2+184320t-31680)x^7+(473088t^6+1419264t^5\\
&+645120t^4-967680t^3-506880t^2+152640t+452480)x^6+(-405504t^7-1419264t^6\\
&-1032192t^5+806400t^4+675840t^3-342720t^2-1708800t-616800)x^5+(253440t^8\\
&+1013760t^7+1075200t^6-161280t^5-345600t^4+532800t^3+2601600t^2+1730400t\\
&+215600)x^4+(-112640t^9-506880t^8-737280t^7-322560t^6-184320t^5-648000t^4\\
&-2163200t^3-1905600t^2-129600t+232800)x^3+(33792t^{10}+168960t^9+322560t^8\\
&+322560t^7+353280t^6+550080t^5+1180800t^4+1195200t^3+324000t^2-36000t+122850)x^2\\
&+(-6144t^{11}-33792t^{10}-81920t^9-126720t^8-184320t^7-267840t^6-449280t^5\\
&-511200t^4-302400t^3-108000t^2-72900t-12150)x+512t^{12}+3072t^{11}+9216t^{10}+19200t^9\\
&+34560t^8+54720t^7+86400t^6+108000t^5+97200t^4+64800t^3+36450t^2+12150t+2025.
\end{aligned}
\end{eqnarray*}

{\bf Acknowledgments} {This work is supported by the NSF of China under Grant No. 11671219 and the K.C. Wong Magna Fund in Ningbo University. K.P. acknowledges DST, NBHM, CSIR, and IFCPAR, Government of India, for the financial support through major projects.}

\begin{figure}[!htbp]
\centering
\subfigure[$s_0=0$, $R_{2}=0.8$]{\includegraphics[height=6cm,width=6cm]{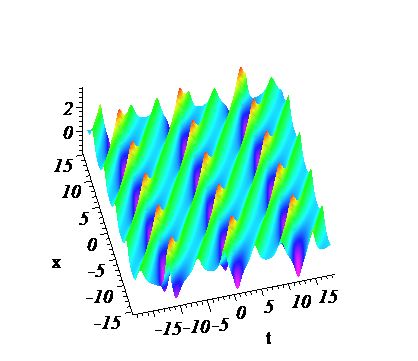}}\qquad\qquad
\subfigure[$s_0=0$, $R_{2}=0.8$]{\includegraphics[height=5cm,width=5cm]{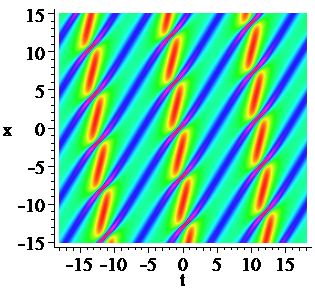}}
\subfigure[$s_0=0$, $R_{2}=0.71$]{\includegraphics[height=6cm,width=6cm]{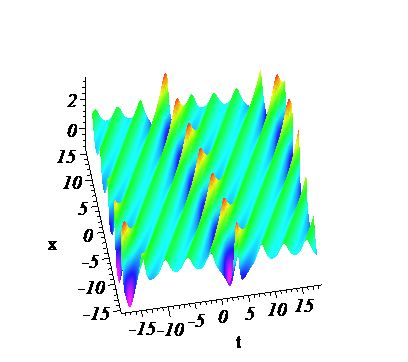}}\qquad\qquad
\subfigure[$s_0=0$, $R_{2}=0.71$]{\includegraphics[height=5cm,width=5cm]{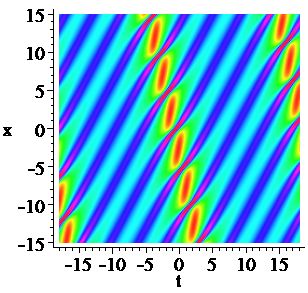}}
\subfigure[$s_0=0$, $R_{2}=0.55$]{\includegraphics[height=6cm,width=6cm]{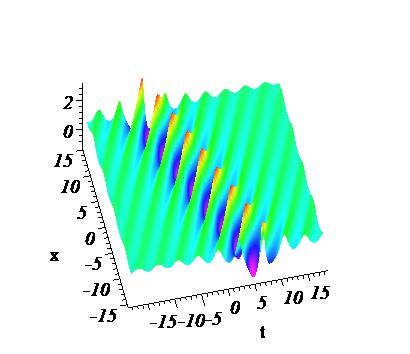}}\qquad\qquad
\subfigure[$s_0=0$, $R_{2}=0.55$]{\includegraphics[height=5cm,width=5cm]{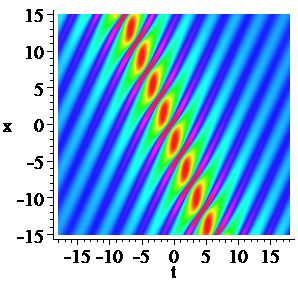}}
\caption{The order-$2$ periodic breather $q^{[2]}_{\rm br}$ gradually approaches the order-$2$ breather-positon
solution when $R_{2}$ goes to $R_{1}$. The right column is the density plot of the left column. The last row is almost the same as the
profile of the order-$2$ breather-positon solution in Fig. \ref{fig.8}. Here $\alpha=-\frac{1}{6}$ and $R_{1}=0.5$. }\label{fig.2}
\end{figure}

\begin{figure}[!htbp]
\centering
\subfigure[$s_0=0$, $R_{1}=0.5$.]{\includegraphics[height=6cm,width=6cm]{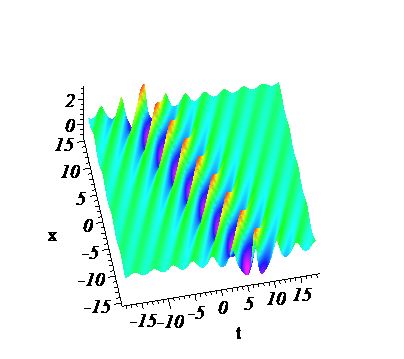}}\qquad\qquad
\subfigure[$s_0=0$, $R_{1}=0.5$.]{\includegraphics[height=5cm,width=5cm]{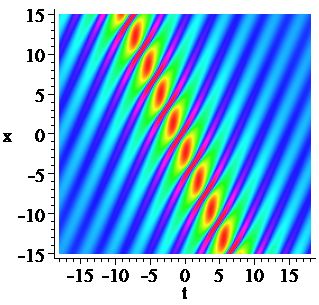}}
\subfigure[$s_0=0$, $R_{1}=0.8$]{\includegraphics[height=6cm,width=6cm]{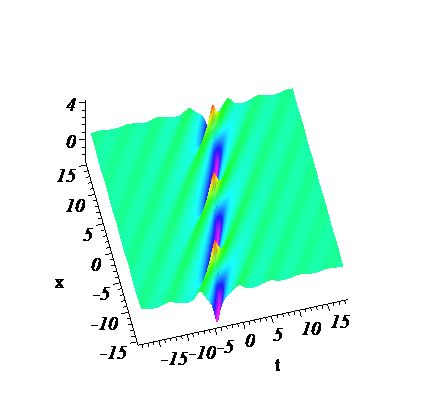}}\qquad\qquad
\subfigure[$s_0=0$, $R_{1}=0.8$]{\includegraphics[height=5cm,width=5cm]{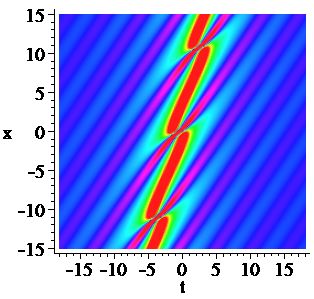}}
\subfigure[$s_0=0$, $R_{1}=0.99$]{\includegraphics[height=6cm,width=6cm]{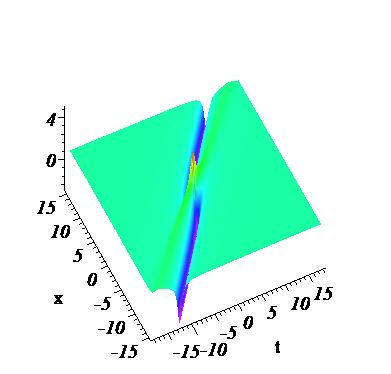}}\qquad\qquad
\subfigure[$s_0=0$, $R_{1}=0.99$]{\includegraphics[height=5cm,width=5cm]{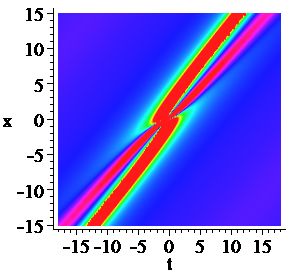}}
\caption{The order-$2$ breather-positon solution gradually approaches the order-$2$ rational solution
 when $R_1$ goes to $1$. The right column is the density plot of the left column. The last row is almost the same as the profile of the order-$2$ rational solution in Fig. \ref{fig.9}.  Here $\alpha=-\frac{1}{6}$ and $s_{1}=0$.}\label{fig.3}
\end{figure}

\begin{figure}[!htbp]
\centering
\subfigure[]{\includegraphics[height=7cm,width=7cm]{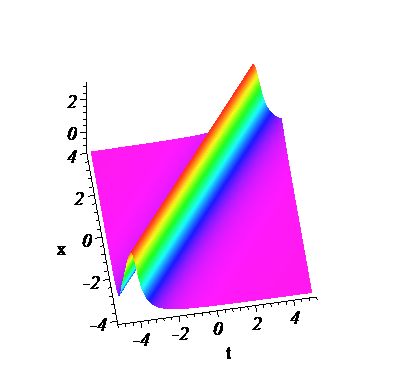}}\qquad\qquad
\subfigure[]{\includegraphics[height=5cm,width=5cm]{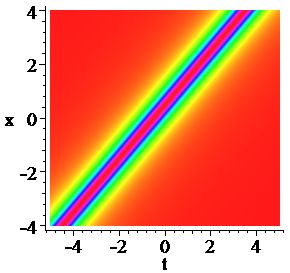}}
\caption{
The order-$1$ rational solution in the $(x,t)$-plane (a) and the density plot (b). Here $\alpha=-\frac{1}{6}$.}\label{fig.4}
\end{figure}

\begin{figure}[!htbp]
\centering
\subfigure[]{\includegraphics[height=7cm,width=7cm]{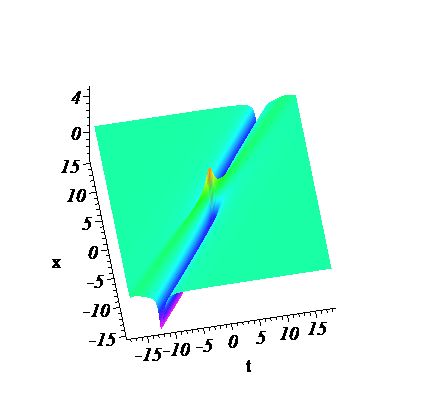}}\qquad\qquad
\subfigure[]{\includegraphics[height=5cm,width=5cm]{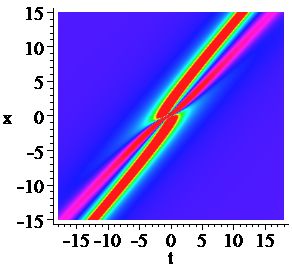}}
\caption{The order-$2$ rational solution in the $(x,t)$-plane (a) and the density plot (b). Here $\alpha=-\frac{1}{6}$. }\label{fig.5}
\end{figure}

\begin{figure}[!htbp]
\centering
\subfigure[]{\includegraphics[height=7cm,width=7cm]{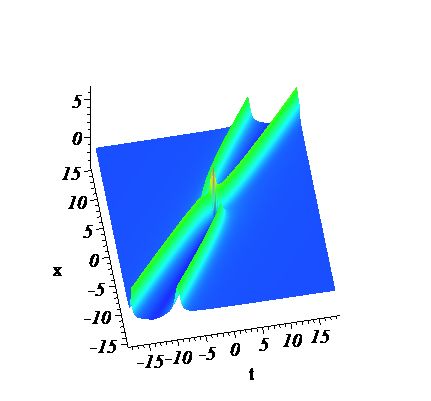}}\qquad\qquad
\subfigure[]{\includegraphics[height=5cm,width=5cm]{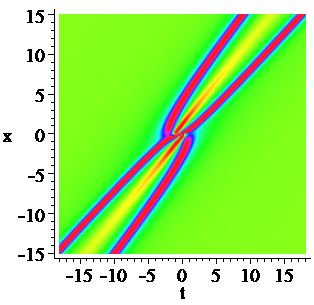}}
\caption{The order-$3$ rational solution in the $(x,t)$-plane (a) and the density plot (b). Here  $\alpha=-\frac{1}{6}$.}\label{fig.6}
\end{figure}

\begin{figure}[!htbp]
\centering
\subfigure[]{\includegraphics[height=7cm,width=7cm]{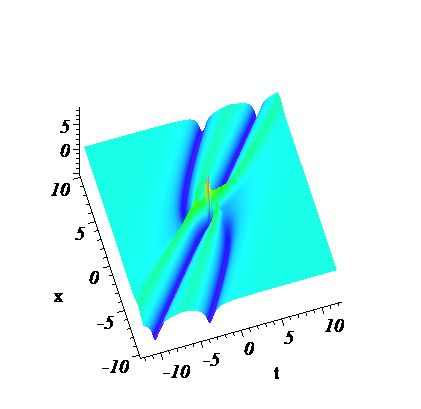}}\qquad\qquad
\subfigure[]{\includegraphics[height=5cm,width=5cm]{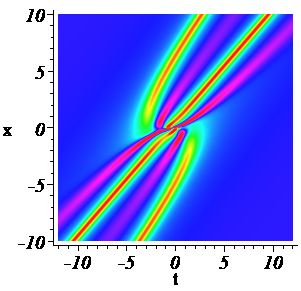}}
\caption{The order-$4$ rational solution in the $(x,t)$-plane (a) and the density plot (b). Here  $\alpha=-\frac{1}{6}$.}\label{fig.7}
\end{figure}

\begin{figure}[!htbp]
\centering
\subfigure[x=-5]{\includegraphics[height=5cm,width=5cm]{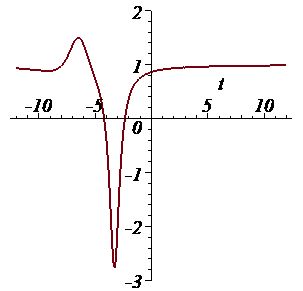}}\qquad
\subfigure[x=-0.128]{\includegraphics[height=5cm,width=5cm]{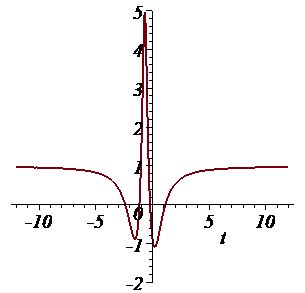}}\qquad
\subfigure[x=5]{\includegraphics[height=5cm,width=5cm]{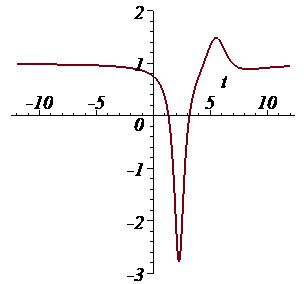}}
\caption{Three pulses of an order-$2$ breather-positon solution $q^{[2]}_{\rm b-p}$ at different locations for $\alpha=-\frac{1}{6}$,  $s_{0}=0$, $s_{1}=0$, and $R_{1}=0.99$. The amplitude is $1.498$ at $t=-6.492$ for panel (a), $4.96$ at $t=-0.628$ for panel (b) and $1.481$ at $t=5.51$ for panel (c). Panel (a) shows the initial pulse, and the other panels show the evolution of the input pulse at two other values of the coordinate $x$.}\label{fig.8}
\end{figure}

\begin{figure}[!htbp]
\centering
\subfigure[x=-5]{\includegraphics[height=5cm,width=5cm]{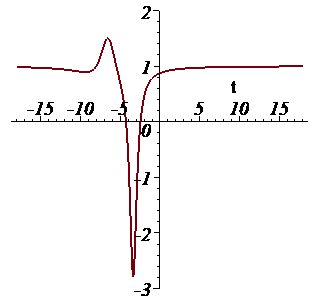}}\qquad
\subfigure[$x=-0.125$]{\includegraphics[height=5cm,width=5cm]{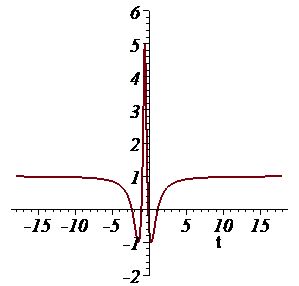}}\qquad
\subfigure[x=5]{\includegraphics[height=5cm,width=5cm]{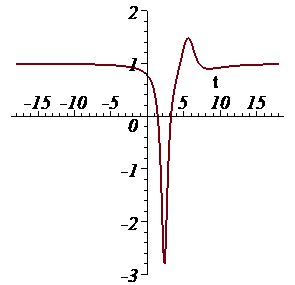}}
\caption{Three pulses of an order-$2$ rational solution $q^{[2]}_{\rm r}$ at different locations for $\alpha=-\frac{1}{6}$, $s_{0}=0$, $s_{1}=0$, and $R_{1}=1$.
The amplitude is $1.501$ at $t=-6.57$ for panel (a), $5$ at $t=-0.625$ for panel (b), and
 $1.484$ at $t=5.57$ for panel (c). Note that
the corresponding pulses at same locations are very similar to each other in Figs. \ref{fig.8} and \ref{fig.9}, which supports strongly the conclusion that the breather-positon solution is indeed an excellent approximation of the corresponding rational solution. }\label{fig.9}
\end{figure}

\end{document}